\documentstyle[11pt]{article}
\begin{document}
\newcommand{\be}{\begin{equation}}
\newcommand{\ee}{\end{equation}}
\newcommand{\bea}{\begin{eqnarray}}
\newcommand{\eea}{\end{eqnarray}}
\newcommand{\beaa}{\begin{eqnarray*}}
\newcommand{\eeaa}{\end{eqnarray*}}
\newcommand{\qd}{\quad}
\newcommand{\qqd}{\qquad}
\newcommand{\npb}{\nopagebreak[1]}
\newcommand{\nn}{\nonumber}
\title{\bf On equation of geodesic deviation and its solutions}
\author{V.S. Dryuma\thanks{Work supported in part by MURST,Italy.
Permanent address: Institute of Mathematics
of Moldavian Academy of Sciences,
Kishinev 277028, Moldova; Academitcheskaya str.5, e-mail: 
15valery@mathem.moldova.su}
\hspace{0.1em} and B.G. Konopelchenko\\
Consortium EINSTEIN
\\ Dipartimento di Fisica dell'Universit\`a
and Sezione INFN, \\73100 Lecce, Italy}
\date{}
\maketitle
\begin{abstract}
Equations of geodesic deviation for the 3-dimensional and 4-dimensional
 Riemann spaces
are discussed. Availability of wide classes of exact 
solutions of such equations, due to recent results for the matrix 
Schr\"odinger 
equation,
is demonstrated. Particular classes of exact solutions for the geodesic
deviation equation as well as for the Raychaudhuri and generalized
Raychaudhuri equation are presented. Solutions of geodesic deviation 
equation for the Schwarzshild and Kasner metrics are found. 
\end{abstract}
\section{Introduction}
     It is well-known that the equation of geodesic deviation is an important
equation in general relativity. It relates the relative acceleration between
two test particles to certain components of the Riemann-curvature tensor.
A knowledge of geodesic deviations is also needed to evaluate or simplify the 
first and higher derivatives of two-point geometrical quantities( the world 
function or geodetic interval), the parallel propagator and others geometric
quantities related to these [1-3].
     
     The Raychaudhuri equation, which is closely connected with the equation 
of geodesic deviation plays a key role in an analysis of the focusing 
effects of gravity [3,4]. Evolution of deformations of a relativistic 
membranes
and the focusing of families of surfaces are described by the 
generalisations of the Raychaudhuri equation [5,6].

     Equation of geodesic deviation is also an important in mathematics, 
in the theory 
of Riemann spaces. For surfaces it coincides, in essence, with the Gauss 
equation in 
geodesic coordinates while in the three and higher-dimensional Riemann spaces
it carries 
an essential information about these spaces. Solutions of the 
geodesic deviations equation (Jacoby fields) and their properties(e.g., an 
existence of conjugate points) are related to various important 
characteristics of Riemann spaces [7-8].

     The equations of geodesic deviation are the system of second order 
linear equations. Their relevance to various problems in physics and 
mathematics
has became clear long time ago. But, suprisingly, still not many exact 
solutions of these equations have been found. They basically correspond to 
the particular case of reduction to a single second order equation [4,6,9].

     In this paper we will present a wide variety of exact soliutions of the 
equation of geodesic deviations. These solutions became available due to 
the recently discovered method of the inverse spectral transform [10,11].
We will consider equations of geodesic deviation in the two, three and four 
dimensional 
Riemann spaces as well as the Raychaudhuri equation and its generalisations.

      The case of the diagonalized separated system of the geodesic deviation 
equations which correspond to the diagonal metric is considered in details.
The nondiagonalizable case is studied with the use of the theory of the matrix
Schr\"odinger equation. General formulae and particular solutions are 
presented.
 
       Specification of the geodesic deviations equation via nonlinear 
integrable (soliton) equations is discussed.
Solutions of the geodesic 
deviations equation in general relativity for
 the Schwarzschild and Kasner 
metrics are presented.

\section{Equation of geodesic deviations}

     Here we will present some known formulas, which will be used in what 
follows.
In the Riemann space with the metric
\begin{equation}
      ds^2 = g_{ij}dx^idx^j
\end{equation}
the geodesics are defined via the equation
\begin{equation}
      \frac{d^2x^i}{ds^2} + \Gamma^i_{kj}\frac{dx^i}{ds}\frac{dx^j}{ds}=0
\end{equation}
where  $\Gamma^i_{kj}$ are the Christoffel symbols of metric.
     
     Let $\eta^i$  be the components of deviation vector between two 
infitesimally nearby geodesic lines. Then these components satisfy to the 
Jacoby equation [1]
\begin{equation}
v^i{\nabla_ i}(v^j{\nabla_ j}\eta^l)=-v^iR^l_{ikm}v^m\eta^k
\end{equation}
where $v^i$ are the components of tangent vector along geodesic line 
$\gamma$, symbol $\nabla_i$ means the covariant derivation along the 
vector field  $v^i$  with respect to the metric (1) and $R^i_{jkl}$ are
the curvature tensor of the metric (1). 
 
     In a special system of coordinates were axis $x^i$ is a geodesic line
equations (3) have the following form [1-3]
\begin{equation}
\frac{d^2\eta^j}{dx^{{i}^2}} + R^j_{ili}\eta^l =0 .
\end{equation}
This system of equations is a main object of our consideration.

     The system of equations (4) has a form of the matrix Schr\"odinger
operator:
\begin{equation}
    [-\frac{d^2}{dx^2} + U(x)]\Psi=\lambda\Psi
\end{equation}
where $U_{ij}(x)=(-R^j_{ili} +\lambda\delta_{jl})$, $x^i=x$ and $\Psi$ is the 
fundamental matrix-solution.

     In the one-dimensional case the equation of geodesic deviation is 
equivalent to 
the Gauss equation 
\begin{equation}
\frac{d^2\Psi}{dx^2} +K(x,y)\Psi=0
\end{equation}
which connects the curvature $K(x,y)$ of the surface with the metric
\begin{equation}
ds^2=dx^2+\Psi^2(x,y)dy^2. 
\end{equation}

     The Raychaudhuri equation also is, in fact, the one-dimensional 
Schr\"odinger equation [4,3]. Indeed, introducing the variable 
$\Theta=\frac{\partial}{\partial x}\log\det U$, one gets from equation (4) 
the following equation
$$
\frac{d\Theta}{dx}=-\frac{1}{3}\Theta^3-\sigma^2+\omega^2-R,
$$
where $\sigma^2=\sigma^{ab}\sigma_{ab}$, $\omega^2=\omega^{ab}\omega_{ab}$
and $R$ are certain quantities depending on the metric. 
The change of variable
$$
\Theta=3\frac{\Psi_x}{\Psi}
$$
leads to the equation
$$
\Psi_{xx}=\frac{1}{3}(\omega^2-\sigma^2-~R)\Psi.
$$
In particular, for the isotropic cosmological model with the Einstein 
equation
$$
 -8\pi T^{\mu}_{\nu}=R^{\mu}_{\nu}-\frac{1}{2}R\delta^{\mu}_{\nu}+
\Lambda\delta^{\mu}_{\nu}
$$
one has
\begin{equation}
\frac{d^2}{dt^2}\Psi 
=\frac{1}{3}[\Lambda-4\pi\rho -\phi^2]\Psi
\end{equation}
where $\rho$ is a density of matter and $\phi$ is a certain geometric 
characteristic.
     
     The Raychaudhuri equation is the principal equation for an analysis
of the singularity behaviour in general relativity. Few particular solutions 
of this 
equation have been found untill now.
      
     Wide classes of solutions of equation(6) have been presented in the 
paper [12] . Since for surfaces  $\eta=\Psi(x,y)dy$  then the results of [12] 
provide wide classes of the Jacoby field for the surfaces.

\section{Geodesic deviation equation  for three-dimensional space}
 
    In the case of three-dimensional space the use of the coordinate 
transformations allows us to reduce a general metrics to the form
\begin{equation}
ds^2=dx^2 +A(x,y,z)dy^2 +2B(x,y,z)dydz +C(x,y,z)dz^2 .
\end{equation}

For such a metric the matrices of the Christoffel's symbols are

\begin{center}
$$
\Gamma_1=\left( \begin{array}{ccc}
0 & 0 & 0 \\
& & \\
0 & \frac{CA_x-BB_x}{2\Delta} & \frac{CB_x-BC_x}{2\Delta} \\
& & \\
0 & \frac{AB_x-BA_x}{2\Delta} & \frac{AC_x-BB_x}{2\Delta}
\end{array}\right),
$$
\end{center}

\begin{center}
$$
\Gamma_2=\left( \begin{array}{ccc}
0 & -\frac{A_x}{2} & -\frac{B_x}{2} \\
& & \\
\frac{CA_x-BB_x}{2\Delta} & \frac{CA_y+BA_z-2BB_y}{2\Delta} & 
\frac{CA_z-BC_y}{2\Delta} \\
& & \\
\frac{AB_x-BA_x}{2\Delta} & \frac{2AB_y-BA_y-AA_z}{2\Delta} & 
\frac{AC_y-BA_z}{2\Delta}
\end{array}\right),
$$
\end{center}

\begin{center}
$$
\Gamma_3=\left( \begin{array}{ccc}
0 &-\frac{B_x}{2} & -\frac{C_x}{2} \\
& & \\
\frac{CB_X-BC_x}{2\Delta} & \frac{CA_z-BC_y}{2\Delta} & 
\frac{2CB_z-CC_y-BC_z}{2\Delta} \\
& & \\
\frac{AC_x-BB_x}{2\Delta} & \frac{AC_y-BA_z}{2\Delta} & 
\frac{AC_z+BC_y-2BB_z}{2\Delta}
\end{array}\right)
$$
\end{center}
where $\Delta= AC-B^2$.

     The equations of geodesic deviations are of the form
\begin{eqnarray}
\frac{d^2\eta^2}{dx^2} + R^2_{121}\eta^2 + R^2_{131}\eta^3 =0,\nn \\
\frac{d^2\eta^3}{dx^2} + R^3_{121}\eta^2 + R^3_{131}\eta^3 =0
\end{eqnarray}
where the corresponding components of the curvature tensor are
$$
R^2_{121}=-\frac{1}{2}\frac{\partial}{\partial x}(\frac{CA_x-BB_x}{\Delta})-
\frac{(CA_x-BB_x)^2}{4\Delta^2}-\frac{(CB_x-BC_x)(AB_x-BA_x)}{4\Delta^2},\\
$$
$$
R^2_{131}=-\frac{1}{2}\frac{\partial}{\partial x}(\frac{CB_x-BC_x}{\Delta})-
\frac{(CB_x-BC_x)\Delta_x}{4\Delta^2},\\
$$
$$
R^3_{121}=-\frac{1}{2}\frac{\partial}{\partial x}(\frac{AB_x-BA_x}{\Delta})-
\frac{(AB_x-BA_x)\Delta_x}{4\Delta^2},\\
$$
$$
R^3_{131}=-\frac{1}{2}\frac{\partial}{\partial x}(\frac{AC_x-BB_x}{\Delta})-
\frac{(AC_x-BB_x)^2}{4\Delta^2}-\frac{(AB_x-BA_x)(CB_x-BC_x)}{4\Delta^2}.
$$

     The system of equations (10) can be rewriten in form of the matrix
Schr\"odinger operator
\begin{eqnarray}
-\frac{d^2\eta^2}{dx^2} + (-R^2_{121} +\lambda^2)\eta^2 +(- 
R^2_{131})\eta^3=\lambda^2\eta^2,\\ \nn
-\frac{d^2\eta^3}{dx^2} + (-R^3_{121})\eta^2 +(- 
R^3_{131} +\lambda^2)\eta^3=\lambda^2\eta^3.
\end{eqnarray}

\section{Some examples of solutions of geodesic deviation equation}
 
     The simplest case of the deviation equations (7) corresponds to the
diagonal metrics, i.e. to the case $B=0$. In this case we have the system
\begin{eqnarray}
-\frac{d^2\eta^2}{dx^2} + \frac{d^2A}{dx^2}\eta^2=0,\\ \nn
-\frac{d^2\eta^3}{dx^2} + \frac{d^2B}{dx^2}\eta^3=0 \nn
\end{eqnarray}
and its solutions correspond to the geodesic deviations for the 3-dimensional
orthogonal metrics (9). We can construct wide classes of solutions of each of 
equations (12)
using the results for the one-dimensional Schr\"odinger equation. In this case
the components of deviations vector $\eta^2$ and $\eta^3$ are completely 
independent.
 
     One can relates them by some special conditions. One possibility is to 
consider the system of equations (12) as related by the Darboux 
transformations,
i.e. 
$$
\frac{d^2B}{dx^2}=\frac{d^2A}{dx^2}-2\frac{\partial^2}{\partial 
x^2}\log\eta^2_1,\\
$$
$$
\eta^3=\eta^2_x -\frac{\eta^2_{1x}}{\eta^2_1}\eta^2
$$
where $\eta^2_1$ is  some solution of the first equation (12).
The set of metrics related by the conditions of such type remains always 
undetermined and for its further specification it is necessary to fix the 
dependence on the variables $y$ and $z$. We can fix 
this dependence with the help of the some equations or to consider the 
metrics 
as induced from the immersion in some Euclidean space or another type of 
spaces. 

     As example we can consider the class of 3-dimensional orthogonal 
metrics
$$
  ds^2=A^2(x,y,z)dx^2+B^2(x,y,z)dy^2+C^2(x,y,z)dz^2
$$
and will require that the functions $A$, $B$, $C$ obey the Darboux equations
[13,14] 
$$
     A_{zy}=\frac{C_y}{C}A_z +\frac{B_z}{B}A_y, \\
$$  
$$  
 B_{zx}=\frac{C_x}{C}B_z +\frac{A_z}{A}B_x, \\ 
$$  
$$  
 C_{xy}=\frac{A_y}{A}C_x +\frac{B_x}{B}C_y. \\ 
$$     
In the case $A=1$  this system is reduced to the relations
$$
              B_z=U(y,z)C, \quad C_y=V(y,z)B,
$$
which are equivalent to the Laplace-type equation for the functions $A$ and 
$B$
$$
B_{zy}-\frac{U_y}{U}B_z-UVB=0, \quad C_{zy}-\frac{V_z}{V}C_y-VUC=0.
$$
These systems of equations are well known in  theory of integrable 
equations and one can use different methods to construct their exact
solutions.           
   
     In any case using the N-times iterated Darboux transformation and 
certain addition conditions one can get the discrete set of metrics
$$  
  ds^2_n=dx^2 +B_n(x,y,z)dy^2 + C_n(x,y,z)dz^2,
$$
which may be of interest from the various points of view.

     For analysis of nondiagonal metrics (9) we can use the results 
on the matrix Schr\"odinger equation and corresponding nonlinear integrable 
differential equations.

     The simplest way consists in the use of the relations between the 
AKNS-system [10,11]
$$
\frac{\partial\psi_1}{\partial x} +i\lambda\psi_1=q(x,y,z)\psi_2,\\
$$      
$$
\frac{\partial\psi_2}{\partial x}-i\lambda\psi_2=r(x,y,z)\psi_1
$$
and the corresponding Schr\"odinger-like system of equations [15]:
$$
\left[-\left(\begin{array}{cc}
1 & 0 \\
0 & 1 \\
\end{array}\right)\frac{\partial^2}{\partial x^2} +
\left(\begin{array}{cc}
rq & q_x \\
r_x & rq\\
\end{array}\right)\right]\left(\begin{array}{c}
\psi_1 \\
\psi_2 \\
\end{array}\right)=\lambda^2\left(\begin{array}{c}
\psi_1 \\
\psi_2 \\
\end{array}\right).
$$
Comparing these equations with  equations (8), one obtains the following
conditions on curvature tensor
$$
\lambda^2-R^2_{121}=rq,\quad    \quad \lambda^2-R^3_{131}=rq, \\
$$
$$
R^2_{131}=-q_x, \quad    \quad R^3_{121}=-r_x.
$$
These conditions imply certain relations between components of
metric. 
     
    In fact, from the condition
$$
R^2_{121} =R^1_{131}
$$
we have
$$
CA_x-AC_x=\delta(y,z)\sqrt{AC-B^2}.
$$
Other two  conditions lead to the relations
$$
 AB_x-BA_x=2\sqrt{AC-B^2}\left(\int r_x\sqrt{AC-B^2}dx +\alpha(y,z)\right),
$$                              
$$
 CB_x-BC_x=2\sqrt{AC-B^2}\left(\int q_x\sqrt{AC-B^2}dx + \beta(y,z)\right)
$$
where $\alpha(y,z)$,\quad $\beta(y,z)$ and 
$\delta(y,z)$ are arbitrary functions.

     So, we have some relations between the functions and we can hope that 
the solutions of this system exist. As result, we can obtain the
two parametrical family of solutions of the matrix Schr\"odinger equation 
and then we can fix  dependence on variables $y$ and $z$ by 
nonlinear integrable equation which have the form of the Lax-equations
$$
L_y=[L,A] \quad or \quad L_z=[L,A]
$$
where
$$
L=-\frac{d^2}{dx^2} +U(x,y,z) 
$$

and $~A$ is some matrix differential operator.

     So this approach to the solutions of geodesic deviation equations 
 allows us to calculate the examples of 3-dimensional metrics which could
be useful in the various problems (for example, in general relativity). 

\section{ Solutions of the matrix Schr\"odinger equation and geodesic
 deviation equation in the 4-dimensional space} 
 
     We see that the construction of solutions of the matrix Schr\"odinger 
equation is a key step in the study of the geodesic deviation equation.
     
     The matrix Schr\"odinger equation has been studied recently within the 
framework of the inverse spectral transform method [15,16]. We will present 
here some results from [16].

     Solution of equation (5) is given by the formulas
$$ 
\Psi = \exp(ikx)+\int_x^{\infty}K(x,x')\exp(ikx')dx'
$$
where $k^2=\lambda$ and  
$$
U(x)=-2\frac{d}{dx}K(x,x).
$$
Here the matrix function $K(x,x')$ is the solution of linear integral 
equation
$$
K(x,x')+M(x+x')+\int_x^{\infty}dx''K(x,x'') M(x'+x'')=0,\quad x < x'
$$
with the kernel
$$
M(x)=\sum_{n=1}^N{C_n\exp(-k_nx)} +\frac{1}{2\pi}\int_{-\infty}^{+\infty} 
d\lambda R(\lambda)\exp(i\lambda x)
$$
where $C_n$ and $R(\lambda)$ are arbitrary matrices.
The simplest solution of this integral equation corresponds to the case
$R=0$ and $N=1$. The corresponding potential $U$ and wave function $\Psi$ are
\begin{equation}
Q(x)=-2p^2\cosh^2[p(x-\xi)]P,
\end{equation}
$$
 \Psi=(ik-p\tanh[p(x-\xi)])\exp(ik(x-\xi))P 
$$
where $P$ is projector, i.e.the matrix which satisfies the condition
$$
P^2=P.
$$
     
      More complicated solution corresponds to $N=2$ and $R=0$. It is of the 
form
$$
U(x,y,z)= -W_x,
$$
$$
W(x,y,z) =-2(p_1+p_2)[1-\rho\tau_1\tau_2]^{-1}
[\tau_1P_1+\tau_2P_2-\tau_1\tau_2\{P_1,P_2\}],
$$
$$
\tau_k=\frac{P_k}{(p_1+p_2)]}(1-\tanh[p_k(x-\xi_k)])
$$
and the function $\rho$ is defined from the relations
$$
\rho P_1=P_1 P_2 P_1,
$$
$$ 
\rho P_2=P_2 P_1 P_2.
$$

     All solutions which correspond to the case $R=0$ can be written 
in a closed 
form [15,16]. These exact solutions could be used to specify the components 
of the 
curvature tensor.
For instance, for the solution (13) one has the following relations between 
components of curvature tensor
$$
R^2_{131}R^3_{131}=R^2_{121}R^3_{121}.
$$

     In the case of 4-dimensional space with the geodesic coordinate system
$$
ds^2=dt^2 +g_{ab}dx^adx^b
$$
the geodesic deviations equation has the form
$$
\frac{d^2}{dt^2}\eta^1 +R^1_{010}\eta^1+R^1_{020}\eta^2+R^1_{030}\eta^3=0,
$$
$$
\frac{d^2}{dt^2}\eta^2 +R^2_{010}\eta^1+R^2_{020}\eta^2+R^2_{030}\eta^3=0,
$$
$$
\frac{d^2}{dt^2}\eta^3 +R^3_{010}\eta^1+R^3_{020}\eta^2+R^3_{030}\eta^3=0
$$
where $R^i_{jkl}$ are the components of curvature tensor.
In this case the conditions on $R^i_{jkl}$ to be projector have  more
complicated form.
    
     Using the  generalizations of Wronskian relations between 
the various solutions $\Psi_i$ of the matrix Schr\"odinger equation one
can get the nonlinear relations for the potentials: $Q(x,y,z)$ and
$Q(x,y+\Delta y, z+\Delta z)$ at the fixed value of variable $x$ such that the
corresponding matrix coefficients  $R(\lambda,y,z,)$ and 
$C(y,z,x) = SS^T$  will satisfy linear equations.
     
     In a such a way the nonlinear "boomeron" integrable equations of the form
$$
Q_y(x,y,z) = \Phi[Q(x,y,z),Q_z(x,y,z)]
$$
have been constructed [16]. Here $\Phi$ is a certain function. 
Their  solutions can be  used for the construction of the solutions of 
geodesic
deviation equation.

     So we see that the soliton structures and equations may be useful in 
the study of the geodesic deviations equations and in the general 
relativity.    
            
\section{Some solutions of geodesic deviation equation in general relativity}

     Let us consider the case of 4-dimensional Riemann space with the 
metric [17]
$$
ds^2=dt^2-A^2(t)dx^2-B^2(t)dy^2-C^2(t)dz^2.
$$
The corresponding geodesic deviation equations are of the form
$$
\frac{d^2}{dt^2}\eta^1-\frac{A_{tt}}{A}\eta^1=0,
$$
$$
\frac{d^2}{dt^2}\eta^2-\frac{B_{tt}}{B}\eta^2=0,
$$
$$
\frac{d^2}{dt^2}\eta^3-\frac{C_{tt}}{C}\eta^3=0.
$$
  
    If the  metric (25) satisfies the Einstein equations, then:
$$
A=t^{p_1},\quad B=t^{p_2},\quad C=t^{p_3}
$$
where
$$
p_1+p_2+p_3=1,\quad p_1^2+p_2^2+p_3^2=1.
$$
It is the Kasner metric. The variables $p_i$  admit the parametrisation:
$$
p_1=\frac{-s}{1+s+s^2}\quad p_2=\frac{s(s+1)}{1+s+s^2}\quad 
p_3=\frac{1+s}{1+s+s^2}.
$$
Hence, the  corresponding  equations are not independent, namely:
$$
\frac{d^2}{dt^2}\eta^1-\frac{s(s+1)^2}{(1+s+s^2)t^2}\eta^1=0,
$$
$$
\frac{d^2}{dt^2}\eta^2+\frac{s(s+1)}{(1+s+s^2)t^2}\eta^2=0,
$$
$$
\frac{d^2}{dt^2}\eta^3+\frac{(1+s)s^2}{(1+s+s^2)t^2}\eta^3=0.
$$
    Solutions of the equation of the type
$$
y''=cx^{-2}y
$$
have the form:
$$
 y=\begin{array}{c}
c_1x^{s+\frac{1}{2}} +c_2x^{\frac{1}{2}-s}, \\
c_1\sqrt x +c_2\sqrt x \log x \\
c_1\sqrt x \cos(s\log x) +c_2\sqrt x\sin(s\log x),
\end{array} 
$$
where $2s=\sqrt|4c+1|$
and for the first solution  $4c+1 > 0$, for the second solution
 $4c+1=0$ and for the third $4c+1 < 0$.
 These solutions provide us the solutions and properties of the geodesic 
deviations for the Kasner metric.    

     To study the properies of geodesic deviation for the Schwarzschild
solution of Einstein equation it is convenient to use the metric in  the 
Lemeitre-form [17]
\begin{eqnarray}
&ds^2=dt^2-\frac{dR^2}{\left[\frac{3}{2r_g}(R-t)\right]^{2/3}} 
-\left[\frac{3}{2}(R-t)\right]^{4/3}
r_g^{2/3}(d\theta^2+\sin^2\theta d\phi^2).&
\end{eqnarray}
The corresponding equations for geodesic deviations are
$$
\frac{d^2}{dt^2}\eta^1 -\frac{4}{9(R-t)^2}\eta^1=0,
$$
$$
\frac{d^2}{dt^2}\eta^2 +\frac{2}{9(R-t)^2}\eta^2=0,
$$
$$
\frac{d^2}{dt^2}\eta^3 +\frac{2}{9(R-t)^2}\eta^3=0.
$$
The solutions of these equations have the following simple form
$$
\eta^1=c_1(R-t)^{4/3} +c_2(R-t)^{-1/3},
$$
$$
\eta^2 =c_3(R-t)^{1/2} +c_4(R-t)^{1/3},
$$
$$
\eta^3 =c_5(R-t)^{1/2} +c_6(R-t)^{1/3}.
$$
These solutions can be useful for the  study the properties of Schwarzshild 
metric.

\section{Solution of the Raychaudhuri equation}

     The Raychaudhuri equation (8) is the one-dimensional scalar 
Schr\"odinger equation with the potential $$u=4\pi\rho+\phi^2$$ and
the cosmological constant $\lambda$ plays a role of energy 
($\lambda=\frac{\Lambda}{3}$)
So, any exact solution of the Schr\"odinger  equation provides us the 
solution of
the Raychaudhuri equation. 
     Wide class  of them is given by the ,so-called, Bargmann potentials 
and their eigenfunctions
$$
U(x,y)=2(\log\det A)_{xx}
$$
and
$$
\Psi=Re(\exp[-i\lambda_0 x]+\sum^N_{n=1}\frac{\det A_n}{\det A}
\frac{\exp[-(\alpha_n(y)+i\lambda_0) x]}{\alpha_n(y)+i\lambda_0})
$$
where  $y$ denotes all independent variables except $x$, $A$ is the $N 
\times N$ 
matrix with elements
\begin{equation}
A_{nk}=\delta_{nk}+\frac{\beta_n(y)}{\alpha_n(y)+\alpha_k(y)}
\exp[-(\alpha_n(y)+\alpha_k(y))x].
\end{equation}

    The matrix elements of the matrix $A_n$ are given by (15) with the 
substitution of the last column by the column 
$$-\beta_n(y)\exp(-\alpha_n(y)x)$$
(n=1,2,3,...N).

In the simplest case $N=1$ one has:
$$
U=\frac{2\alpha^2(y)}{\cosh^2[\alpha(y)x-\gamma(y)]}-\frac{\Lambda}{3}
$$
$$
\Psi=Re(\frac{i\lambda_0+\alpha(y)\tanh[\alpha(y)x-\gamma(y)]}
{i\lambda_0+\alpha(y)}\exp[-i\lambda_0 x])
$$
where $$\gamma(y)=\frac{1}{2}\log\frac{\beta(y)}{2\alpha(y)}$$ and
$\alpha(y)$, $\beta(y)$ are arbitrary functions.
These solutions have  very special properties: they are transparent for all 
energies, the corresponding function $\Psi$ have in general $N-1$ zeros.

    The dependence of the functions $\alpha$, $\beta$ can be fixed by the 
requirement that 
$U$ and $\Psi$ obey additional equations. Theory of soliton equations
provides us such equations. They are the famous, Korteweg-de Vries (KdV) 
equation and its higher partners.
The KdV equation has the form
$$
U_y +6UU_x +U_{xxx}=0
$$
while $\Psi$ obeys the equation
$$
\Psi_y+4\Psi_{xxx} -6U\Psi_x -3U_x\Psi=0.
$$
The formulae (20) with $\alpha_n=const$ , $\beta_n=\beta_{n0}\exp( 
8\alpha_n^3 y)$
give us the multi-soliton solutions of the KdV equation. These solutions 
have a number of interesting properties.

    The Raychaudhuri equation inherits all properties of these exact
solutions. 
     We will discuss the properties of such soliton like solutions
within the general relativity elsewhere.

\section{Solutions of the generalized Raychaudhuri equation}

     A generalisation of Raychaudhuri equation in the case of the 
two-dimensional
time-like surfaces embedded in a four-dimensional background  has been
derived  in [5]. It has the form
$$
\Delta\gamma +\frac{1}{2}\partial_a\gamma\partial^a\gamma +(M^2)^i_i=0,
$$                                                   
where $\Delta =\nabla^a\nabla_a$, $\nabla_a$ is the world-sheet covariant
derivative and $\partial_a\gamma=\Theta_a$.
    The quantity $(M^2)^i_i$  is connected with the geometric characteristic
of background and  geometry of membrane embedded in this background.

     In a simplest case this equation looks like [6]
$$ 
-\frac{\partial^2 \Psi}{\partial \tau^2}+ 
\frac{\partial^2 \Psi}{\partial \sigma^2}+
\Omega^2(\sigma,\tau)(M^2)^i_i(\sigma,\tau)\Psi=0
$$
where $\Omega^2$ is the conformal factor of the induced metric written in 
isothermal coordinates. 

It is is the second-order linear, hyperbolic, partial differential equation 
for $\Psi$ describing the deformation of the surfaces with respect 
to parameter $\tau$. It coincides with  deviations equation 
in the case when $\Psi$ does not dependent on $\tau$.

     We present this equation in form
\begin{equation}
-(\partial^2_{\tau} -\partial^2_{\sigma})\Psi + U(\tau,\sigma)\Psi=E\Psi
\end{equation} 
where $$U=\Omega^2(M^2)^i_i+E.$$
Particular  solution of this equation has been obtained in [7] by separation 
of variables.
    A wide class of solutions of equation (16) can be constructed by the 
inverse spectral transform method. The simplest from them are of the form 
[18,19].
 For 
$E\neq0$ 
$$
U(\tau,\sigma)=-2E 
+\frac{E(\alpha-\beta)}{\alpha\beta\cosh^2[(\alpha-\beta)(\tau+\sigma)
+\frac{E(\tau-\sigma)}{\alpha\beta}+\gamma]/2},
$$
$$
\Psi(\tau,\sigma)=\frac{\cosh[(\alpha+\beta)(\sigma+\tau+
\frac{E(\tau-\sigma)}
{\alpha\beta}+\delta)]/2}
{\cosh[(\alpha-\beta)(\sigma+\tau+\frac{E(\tau-\sigma)}
{\alpha\beta}+\gamma)]/2}
$$
where $\alpha$, $\beta$, $\gamma$, $\delta$ are arbitrary parameters.

     For the case $E=0$ we have the solution
$$
U(\tau,\sigma) 
=\frac{4\alpha\beta}{\cosh^2[\alpha(\tau+\sigma)+\beta(\tau-\sigma)]}
$$
and
$$
\Psi(\tau,\sigma)=A\frac{\cosh[\alpha(\tau+\sigma)-\beta(\tau-\sigma)+\delta]}
{\cosh[\alpha(\tau+\sigma)+\beta(\tau-\sigma)+\gamma]}.
$$
A very simple solution with $E=0$ is of the form
$$
U(\tau,\sigma)=\frac{6(\tau+\sigma-\alpha(\tau-\sigma))^2-2\beta^2}
{(\tau+\sigma-\alpha(\tau-\sigma)^2+\beta^2)^2}
$$
and
$$
\Psi(\tau,\sigma)=\frac{1}{(\tau+\sigma-\alpha(\tau-\sigma))^2+\beta^2}
$$
with arbitrary parameters $\alpha$, $\beta$.
Such exact solutions of could be important for an analysis of 
dynamics of surfaces.

\section*{Acknowledgement}

     The first author is grateful to the Physics Departement of Lecce 
University
for the financial support and hospitality.

\newpage
\section{References}
\begin{itemize}
\item[1]{C.W.Misner, K.S.Thorne and J.A.Wheeler, Gravitation (Freeman,
San-Francisco, 1973).}
\item[2]{J.L. Synge, Relativity: The General Theory (North-Holland, 
Amsterdam, 1960).}
\item[3]{S.W.Hawking and G.F.R.Ellis, The large 
scale structure of space-time,
Cambridge University Press, 1973.}
\item[4]{A. Raychaudhuri, Phys. Rev., {\bf 98}, 4,(1955).}
\item[5]{R.Capovilla and J.Guven, Phys. Rev. D {\bf 52}, 1072 (1995).}
\item[6]{S.Kar, Phys. Rev. D {\bf 53}, 2071, (1996).}
\item[7]{ M.P.do Carmo, Riemannian geometry, Boston, 1992.}
\item[8]{S.Kobayashi and K.Nomidzu, Foundation of differential geometry, 
v.2, New-York, 1969.}
\item[9]{P.Dolan, P.Choudhury and Y.Safko, J. Austral. Math Soc. {\bf 22}, 
28,(1980).}
\item[10]{V.E.Zakharov, S.V.Manakov, S.P.Novikov,L.P.Pitaevsky, Theory 
of Solitons, Nauka, 1980.}
\item[11]{M.Ablowitz and H.Segur, Solitons and inverse 
scattering transform. SIAM, Philadelphia, 1981.}
\item[12]{B.G.Konopelchenko, Acta Applicandae Mathematicae, {\bf 39}: 
379-387, 
(1995).}
\item[13]{G.Darboux, Lecons sur la theorie des surfaces, t. 1-4, Paris, 
Gauthie-Villars, (1887-1896).}
\item[14]{V.S.Dryuma, Teor.Math.Fyz, {\bf 99}, 241, (1994).}
\item[15]{M.Wadati, T.Kamijo,  Progr.Theor.Phys.,{\bf 52}, 397, (1974).}
\item[16]{F.Calogero, A.Degasperis, Nuovo Cimento  {\bf 32B}, 201 (1976).}
\item[17]{L.Landau and E.Lifshiz, Field Theory, Moscow, 1967.}
\item[18]{V.Dubrovsky and B.Konopelchenko, Inverse Problem, {\bf 9}, 391, 
(1993).}
\item[19]{B.Konopelchenko, Solitons in multidimensions, World 
Scientific, Singapore, 1993.}
\end{itemize}
\end{document}